\documentclass[a4paper,useAMS,fleqn,usenatbib]{mnras}

\usepackage{newtxtext,newtxmath}
\usepackage[T1]{fontenc}
\usepackage{ae,aecompl}

\usepackage{graphicx}	
\usepackage{amsmath}	
\usepackage{amssymb}	
\hypersetup{draft}

\title[A Cold Stellar Stream in Pegasus]{A Cold Stellar Stream in Pegasus}

\author[H D. Perottoni  et al.]{H\'elio D. Perottoni,$^{1,2}$\thanks{E-mail: hperottoni@gmail.com}
Charles Martin$^{2}$,
Heidi Jo Newberg$^{2}$,
Helio J. Rocha-Pinto$^{1}$,\newauthor
Felipe de Almeida-Fernandes$^{1}$,
and Altair R. Gomes-J\'unior$^{1,3}$
\\
$^{1}$Universidade Federal do Rio de Janeiro, Observat\'orio do Valongo,  Lad. do Pedro Ant\^onio 43, 20080-090 Rio de Janeiro, Brazil\\
$^{2}$Department of Physics, Applied Physics and Astronomy, Rensselaer Polytechnic Institute, Tory, NY 12180, USA\\
$^{3}$UNESP - S\~ao Paulo State University, Grupo de Din\^amica Orbital e Planetologia, CEP 12516-410, Guaratinguet\'a, SP 12516-410, Brazil
}

\date{Accepted XXX. Received YYY; in original form ZZZ}

\pubyear{2018}

\begin{document}
\label{firstpage}
\pagerange{\pageref{firstpage}--\pageref{lastpage}}
\maketitle

\begin{abstract}
We report the serendipitous discovery of a stellar stream in the constellation Pegasus in the south Galactic hemisphere. The stellar stream was detected using the SDSS Data Release 14 by means of a matched filter in the color--magnitude diagram that is optimised for a stellar population that is 8 Gyr old with [Fe/H] = $-$0.46 dex, and located at heliocentric distance of 18 kpc. The candidate stream is faint (turnoff point at $r_0 \sim$ 19.6), sparse and barely visible in SDSS photometry.  It is also detected in the (shallower) Pan-STARRs data. The residual stellar density in the $(u-g)_0$, $(g-r)_0$ color--color diagram gives the same estimate for the age and [Fe/H] of this stellar population. The stream is located at a Galactic coordinates $(l,b) = (79.\degr4,-24.\degr6)$ and extends over 9$^\circ$ (2.5 kpc), with a width of 112 pc.  The narrow width suggests a globular cluster progenitor.

\end{abstract}




\section{Introduction}

We are in an era of rapidly increasing knowledge of the complexity of our galaxy, including the discovery of dozens streams of stars ripped from Milky Way satellites in the stellar halo \citep{Newberg2016} and complex substructure of the disk \citep{Gaiadisk18}.  It has been suggested that subhalos (including infalling dwarf galaxies) interact with the disk, to create warps, rings and even the spiral structure \citep{Purcell2011, Laporte2018a, Laporte2018b, Erkal2018}; alternatively, the interaction of the disk with the halo could deform the disk \citep{chequers2017}.  The substructure of the Milky Way teaches about the formation processes that built the halo, and puffed up and warped the disk \citep{Velazquez1999}.

The Sloan Digital Sky Survey \citep[SDSS;][]{york2000} has provided influential discoveries of density substructure in both the stellar halo and the disk, including the discovery of the tidal tails of the Pal 5 globular cluster \citep{odenkirchen2001}, the discovery of substructure in the stellar halo \citep{newberg2002}, the discovery of the cold stellar stream GD1 \citep{GD1}, the Field of Streams image of the major stellar substructure in the north Galactic cap \citep{belokurov2006}, an estimate of the halo mass from the velocities of blue horizontal branch stars \citep{xue2008}, a map of the stellar density of the halo and discs using photometric parallax \citep{juric2008}, and the discovery of vertical displacements of the Milky Way disk \citep{xu2015}. There were also many notable discoveries in other surveys, for example mapping the Sgr dwarf tidal stream in 2MASS \citep{majewski2003}, and mapping the Virgo Overdensity/Virgo Stellar Stream with RR Lyrae stars \citep{vivas2006, sesar2013}.  In recent years other photometric surveys have been used to discover substructure; notably eight dwarf galaxies \citep{DESdwarfs2015} and eleven tidal streams \citep{shipp2018} have been identified in the Dark Energy Survey \citep[DES;][]{DES2005}.  Currently, data from DR2 of the ESA Gaia Satellite \citep{GaiaDR2} is transforming our knowledge of the Milky Way by providing highly accurate proper motions for Milky Way stars.  Recent results that illustrate the extent to which these observations will revolutionize our knowledge of Milky Way substructure include: mapping the velocity substructure of the disk in exquisite detail \citep{Gaiadisk18}, identifying local halo streams in velocity space \citep{Myeong2018} and suggesting the identification of a major merger remnant in the stellar halo that could have created the Milky Way's thick disk \citep{Helmi2018}.

One important goal of the study of Milky Way substructure is to constrain the density distribution of dark matter.  This has been attempted through fitting simulations of dwarf galaxy tidal disruption to observed tidal streams \citep{Willett2009, koposov10, Law2010, Newberg2010, Penarrubia2012, Kupper2015, Bovy2016, Sanderson2017, Dai2018, Bonaca2018}, and the width or gaps in tidal streams \citep{Carlberg2012, Erkal2015, Bovy2016b, Erkal2016, Sandford2017, Koppelman2018}.  Cold tidal streams are particularly useful for constraining the shape of the halo and for measuring its lumpiness; the streams from less massive substructures such as globular clusters are less affected by the properties of the progenitor itself and passing subhalos leave larger gaps in these thinner streams with smaller velocity dispersion.  For a review see \citet{Johnston2016}.  It is possible that the substructure of the disk could be used to constrain the lumpiness or shape of the dark matter halo \citep[Galactoseismology;][]{Chakrabarti2011, Widrow2012, Gomez2016, Chequers2018, Laporte2018a}.

\begin{table*}
\caption{Sample selections}
\label{tab:selections}
\begin{tabular}{lcc}
\hline
\hline
 Sample                  & Description              & Figure       \\
\hline\hline
 SDSSprimary               & Full sample\footnotemark[1]  &  Fig. \ref{fig:CMD_SDSS_stars_1}  \\
 SDSSgr                  &  SSP selection on ($(g-r)_0$, $r_0$)\footnotemark[1] & Fig. \ref{fig:CMD_SDSS_stars_1}; Figs. \ref{fig:scatter_projection} A and B  \\
 SDSSgrgi  &  SSP selection on ($(g-r)_0$, $r_0$) and ($(g-i)_0$, $i_0$)\footnotemark[1]  & Figs. \ref{fig:scatter_projection}C and D; Fig. \ref{fig:stream_on_off}; Fig. \ref{fig:stream_hist}A\\
   SDSSdwarfs                  &  ($1.2 < (g-r)_0 < 1.4$
and $22.3 < r_0 < 22.7$)\footnotemark[1]  & Fig. \ref{fig:stream_hist}B \\
 Pan-STARRsgi   & SSP selection on ($(g-i)_0$, $i_0$)\footnotemark[1] $i_0 < 21.4$ & Fig. \ref{fig:stream_hist}C \\
  SDSSgi\_shallow                     & SSP selection on ($(g-i)_0$, $i_0$) $i_0 < 21.4$& Fig. \ref{fig:stream_hist}D \\
 OnStream Field                     & SDSSprimary ; 78.9$^{\circ} < l' < 79.9^{\circ}$ and $-19^{\circ} > b' > -27^{\circ}$ & Fig. \ref{fig:stream_cmd}A, C, and D. \\
 OffStream Left Field                    & SDSSprimary ; 76.9$^{\circ} < l' < 77.9^{\circ}$ and $-19^{\circ} > b' > -27^{\circ}$  & Fig. \ref{fig:stream_cmd}B, C, and D.  \\
 OffStream Right Field                    & SDSSprimary ; 80.9$^{\circ} < l' < 81.9^{\circ}$ and $-19^{\circ} > b' > -27^{\circ}$  & Fig. \ref{fig:stream_cmd}B, C, and D.  \\
\hline
\end{tabular} \\
\footnotemark[1] Covering 50$^{\circ} < l < 160^{\circ}$ and -10$^{\circ} > b > -60^{\circ}$
\end{table*}

In this paper we present the serendipitous discovery of a cold stellar tidal stream in the Pegasus constellation. The stream was found while using a matched filter likelihood sample selection in color--magnitude space to study the Triangulum-Andromeda overdensity (\citealt{maj04};\citealt{rp04}). In Section 2, we describe the selection technique used to obtain the sample. In Section 3, we present the properties and show the signature of this stellar stream. In Section 4, we summarize the conclusions.

\section{Data and sample selection}

The main stellar sample used in this paper was selected from the Sloan Digital Sky Survey (SDSS DR14; \citealt{Abolfathi}). It consists of 
all objects classified as stars and flagged as clean photometry using the flag "clean=1". The $ugriz$ magnitudes (\citealt{fukugita96}; \citealt{smith02}; \citealt{doi10}) were corrected for Galactic extinction using the dust maps of \citet{schlegel98} as calibrated by \citet{SF11}. We used SDSS photometric data covering 50$^{\circ} < l < 160^{\circ}$ and $-10^{\circ} > b > -60^{\circ}$. It is the same main sample used to map the TriAnd overdensity \citep{perottoni18}.  Hereafter, the sample will be called SDSSprimary.

\begin{figure}
\includegraphics[width=\columnwidth]{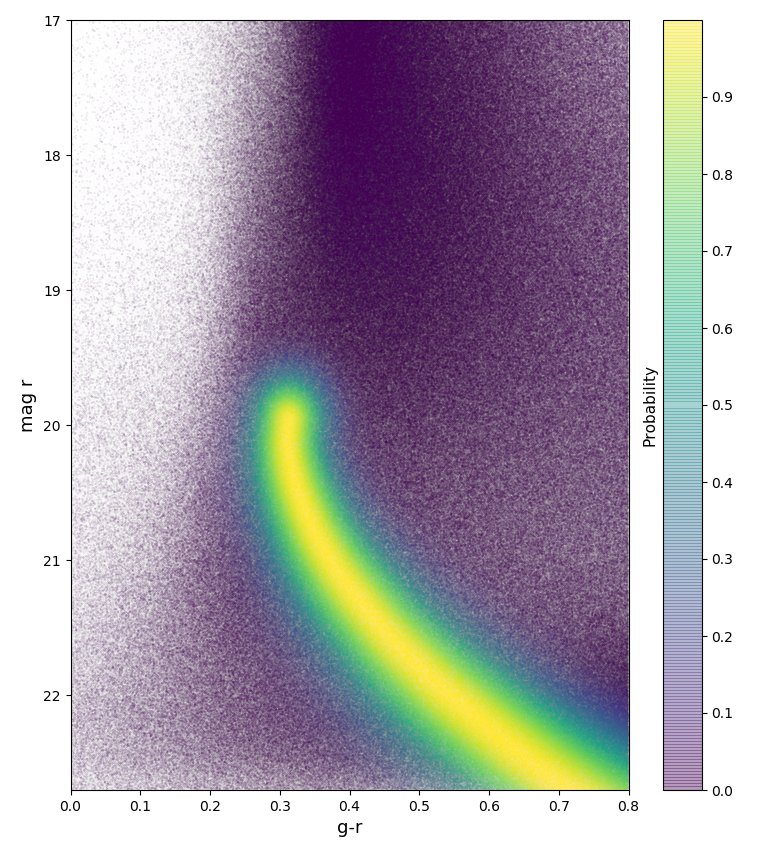}
    \caption{Color--magnitude diagram of SDSS stars located between 50$^{\circ} < l < 160^{\circ}$ and 
$-10^{\circ} > b > -60^{\circ}$. The color scale denotes the probability of a star being a member 
of a 8 Gyr-old single stellar population with a metallicity of [Fe/H] = $-$0.46 and with a distance modulus of 16.3 ($18$ kpc).}
    \label{fig:CMD_SDSS_stars_1}
\end{figure}

The stream was discovered in a plot of celestial coordinates of stars sampled with a high probability of belonging to a single stellar population.
The probability is calculated using the method similar to that described by \citet{JL05}, but instead of estimating a probability for a stellar age we adapted the method to obtain the probability for a set of parameters: age ($\tau$), metallicity  ($\zeta$), and distance ($d$); which represent a simple stellar population (SSP). Equation 3 from \citet{JL05} was rewritten to include the distance:

\begin{equation}
f(\tau, \zeta, m, d) \propto f_0(\tau, \zeta, m, d) \, L(\tau, \zeta, m, d)
\end{equation}
where $f(\tau, \zeta, m, d)\,\mathrm{d}\tau\,\mathrm{d}\zeta\,\mathrm{d}m\,\mathrm{d}d$ is defined as the probability for the star to have an age between $\tau$ and $\tau + \mathrm{d}\tau$, metallicity between $\zeta$ and $\zeta + \mathrm{d}\zeta$, mass between $m$ and $m + \mathrm{d}m$ and distance between $d$ and $d + \mathrm{d}d$. $f_0(\tau, \zeta, m, d)$ is the prior distribution and $L(\tau, \zeta, m, d)$ is the likelihood obtained from models and observations, as described by \citet{JL05}. In our case, we use as observables a color (i.e. g$-$r) and a magnitude (i.e. $r$) and their respective errors.

In order to focus on the probability for a star to belong to a SSP, we marginalize this distribution in mass, obtaining:

\begin{equation}
f(\tau, \zeta, d) = f(SSP) \propto \int f(\tau, \zeta, m, d) \, \mathrm{d}m
\end{equation}
which is discretized as
\begin{equation}
f(SSP) \propto \sum_j f(\tau, \zeta, m_j, d) \, (m_{j+1} - m_{j-1})
\end{equation}

Figure \ref{fig:CMD_SDSS_stars_1} shows a  color--magnitude diagram (CMD) for the $r_0$ vs. $(g-r)_0$  matched filter obtained for the SDSS sample. The color bar indicates the probability that an individual star belonging to a single stellar population that has a metallicity of [Fe/H] = $-0.46$, an age of 8 Gyr, and is located at a distance of 18 kpc. This isochrone comes from the PARSEC database (\citealt{bressan12}). When defining the isochrone matched filter, we decided to truncate the isochrone at its turnoff point ($r_0 \sim$ 19.8) to avoid contamination from nearby disk stars. Although the original isochrone continues towards the red giant branch, we expect that the number of giants stars from the candidate stream should be orders of magnitude smaller than the number of nearby disk stars and no real gain would be possible by using the isochrone filter brighter than the turnoff. The selection was further truncated at $(g-r)_0 \sim 0.8$ to avoid contamination from disk red dwarfs. 

The sample SDSSgr are the stars from SDSSprimary with a probability greater than $0.8$ of belonging to the isochronal filter, as shown in Figure \ref{fig:CMD_SDSS_stars_1}, and will be used to select likely members of the  substructure throughout the rest of this paper. A few additional samples were built to study the candidate stream, and will be used in the upcoming sections. 

Sample SDSSgrgi corresponds to a double isochrone filter, containing stars compatible with the same single stellar population in the $i_0$ vs. ($g-i)_0$ and $r_0$ vs. ($g-r)_0$ CMDs, with a probability greater than $0.8$ in each, as described by \citet{perottoni18}.

To rule out the possibility of contamination by a local structure, we selected a sample of faint SDSS red dwarfs having $1.2 < (g-r)_0 < 1.4$ and $22.3 < r_0 < 22.7$ (hereafter, SDSSdwarfs). These stars are as faint as some of the candidate stream stars and could show us whether we are seeing a real feature or an artifact caused by variation in the depth of the SDSS stripes. 

Sample Pan-STARRsgi was selected as a validation sample covering 50$^{\circ} < l < 160^{\circ}$ and $-10^{\circ} > b > -60^{\circ}$ from the Pan-STARRs DR1 (\citealt{kaiser10}; \citealt{chambers2016}) using the star--galaxy separation $i_{PSF} -i_{Kron} < 0.05$. The Pan-STARRs $g_{P1}i_{P1}$ magnitudes (\citealt{tonry12}) were corrected for Galactic extinction.  The Pan-STARRS sample was selected using a matched filter in the extinction-corrected $g_0$ and $(g-i)_0$ color-magnitude diagram.

All of these samples, as well others explained later, are summarized 
in Table \ref{tab:selections} for easy reference.

\begin{figure*}
\includegraphics[width=\hsize]{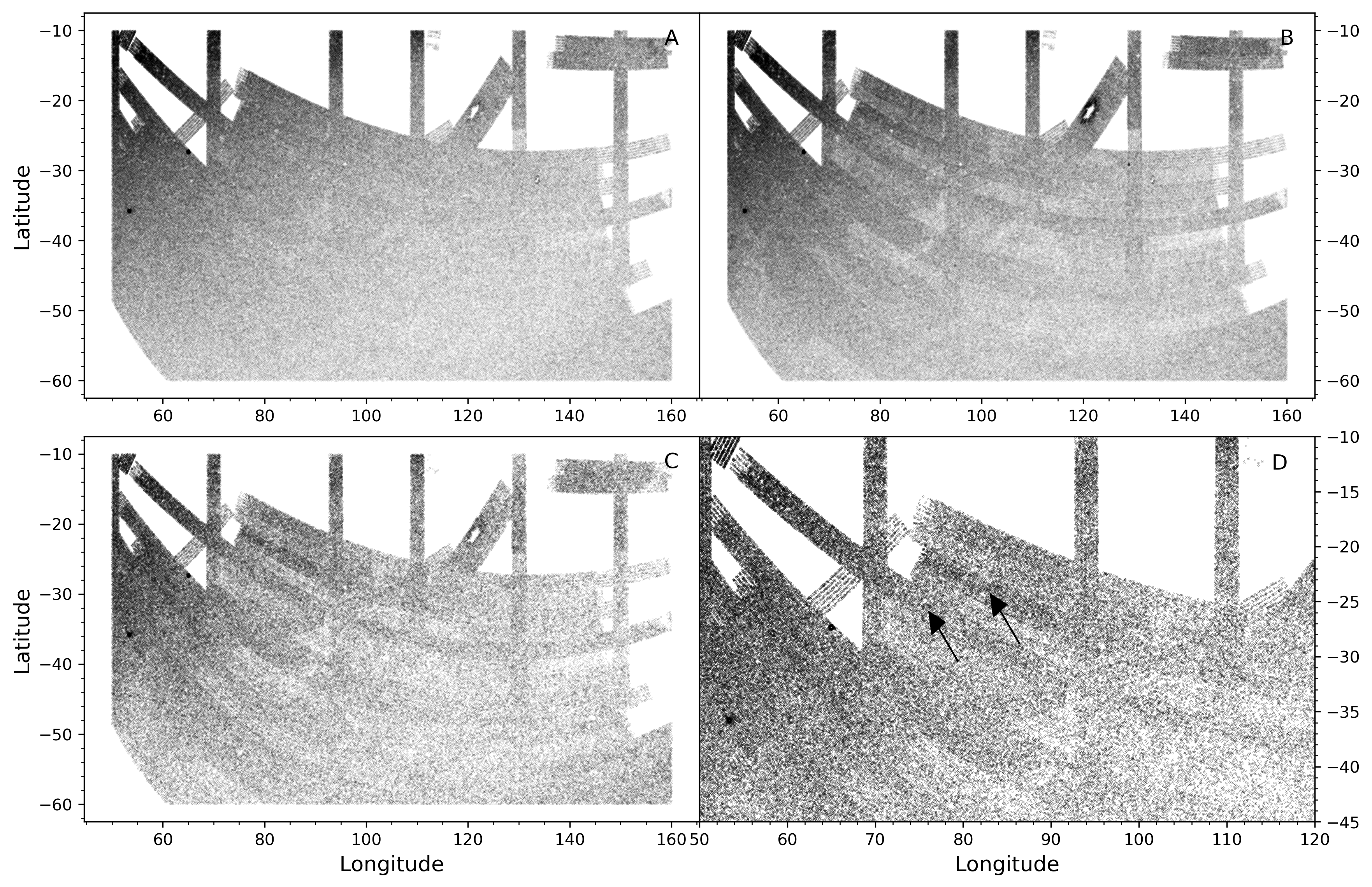}
\caption[density_stelar_map_residual]{Panel A: stellar projection of sample selected to have a probability $> 0.8$ 
in $(g-r)_0$ versus $r_0$, to the magnitude limit of $r_0 < 22.3$, belonging to the SSP (shown in Figure 
\ref{fig:CMD_SDSS_stars_1}). Near $(l,b) = (80\degr,-25\degr)$, a faint stream of stars can be seen cutting across 
the SDSS footprint. Panel B: stellar projection of a sample selection equal to Panel A, but at magnitude limit 
of $r_0 < 22.7$. The stream can be seen more clearly in this panel. Panel C: stellar projection of sample selection 
that is simultaneously compatible with the SSP in two different CMDs ($(g-i)_0$ versus $i_0$ and $(g-r)_0$ versus $r_0$) with a probability $> 0.8$ on both CMDs and to the magnitude limit of $r_0 < 22.7$. Panel D: shows the same projection as Panel C and is zoomed in with the addition of arrows to highlight the stream.}
   \label{fig:scatter_projection}
\end{figure*}

\section{Stream Properties and Significance}

We used the samples defined in the previous section to generate the maps in the region of the candidate stellar stream. Figure \ref{fig:scatter_projection}A shows the scatter map obtained using the SDSSgr sample to a magnitude limit of $r_0 = 22.3$. A faint, narrow stream can be seen at ($l$, $b$) $\sim (79.4^{\circ}, -24.6^{\circ}$) cutting across the corner of the SDSS footprint. This structure seems to point towards the globular M15 located at (l,b) = (65.01, -27.31). However the metallicity and the age from the used SSP is very different from the M15 population. Its angular size and heliocentric distance suggest a structure that measures $\sim 2530$ pc in length and $\sim 112$ pc in width; these values are similar to other faint Galactic streams reported in the literature (\citealt{shipp2018}). See Table \ref{tab:prop} for a summarized list of the stream parameters.

In order to test whether the stream could be found at different distances, we also used the likelihood filter for the same population, shifted to 15 and 21 kpc. The signature of a stream-like candidate cannot be seen at these distances. From this and Fig. \ref{fig:scatter_projection}, we conclude that this stellar stream is located at around 18 kpc from the Sun. We note, however, that matched filter fits to color--magnitude diagrams are known to have large uncertainties in the distance of the stream; for example the distances in \citet{martin13} compared to estimates of distance for Pisces stream with previous works (\citealt{bonaca12}, \citealt{grillmair12}).

Figure \ref{fig:scatter_projection}B differs from Figure \ref{fig:scatter_projection}A by going deeper, down to $r_0 = 22.7$. In this deeper sample, the stream shows a better contrast with respect to the field, although it also highlights the varying completeness of SDSS data at different sky positions using the filter. Because we are able to see the stream in Figure \ref{fig:scatter_projection}A, for which the SDSS $r_0$ magnitude completeness is about $95\%$ for stars \citep{Stoughton2002}, and because the stream does not appear to follow the SDSS stripe pattern (it crosses three stripes), it is unlikely that the stream is a result of varying completeness levels in SDSS.

Figure \ref{fig:scatter_projection}C shows the projection of the stars selected using the sample SDSSgrgi to a magnitude limit of $r_0 < 22.7$. The contrast of the stream is more clear at this selection than in the other two. Figure \ref{fig:scatter_projection}D shows the same selection with arrows to highlight the stream. Since the signal in this panel is clearer, we decided to use this sample to analyze the candidate stream in the remainder of paper.

\begin{table}
\centering
\caption{Parameters of the stellar stream}
\label{tab:prop}
\begin{tabular}{lc}
\hline
\hline
 Parameter               & Value                     \\
\hline\hline
 R.A.                    &  (328.\degr3; 333.\degr4)  \\
 Dec.                    &  (20.\degr8; 28.\degr1)    \\
 $l$                     & (76.\degr4; 85.\degr4)     \\
 $b$                     & ($-$25.\degr5; $-$23.\degr0)   \\
 Age                     & 8 Gyr\footnotemark[1]                      \\
 {\rm [Fe/H]}            & $-$0.46\footnotemark[1]                    \\
$(m-M)_0$                & 16.3 $\pm$ 0.25            \\
 Heliocentric distance   & 18$ \pm $2 kpc              \\
 Galactocentric distance & 18.4 $\pm$ 2 kpc\footnotemark[2]    \\
 Width (FWHM)            & $\sim 0.\degr4$ ($\sim 112$ pc)    \\
 Length                  & $\sim 9.\degr04 $ ($\sim 2530$ pc)                 \\
 Luminosity              & $1-3 \times 10^3 L_{\sun}$             \\
\hline
\end{tabular} \\
\footnotemark[1] Values adopted from the isochronal likelihood filter. \\
\footnotemark[2] We adopted 8.0 kpc for the Solar Galactocentric radius.
\end{table}

\begin{figure}
\includegraphics[width=\columnwidth]{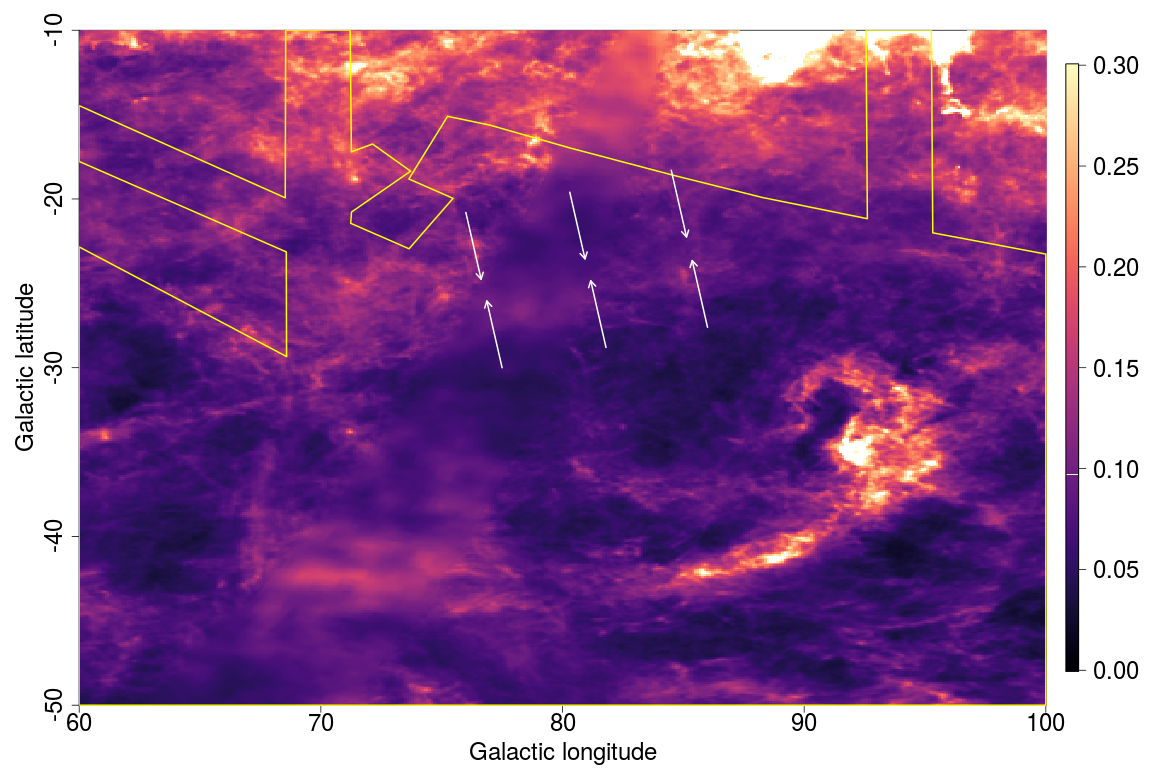}
    \caption{Map of the distribution of the Galactic reddening with $0.^{\circ}1$ resolution. The $E(B-V)$ values in this map come from \citet{schlegel98}. We have found no correlation between lower and higher values of extinction and the stellar stream reported in this paper. The white arrows point to the position of the stream and the yellow line shows the sky coverage of SDSS survey in the region of the stream.}
    \label{fig:extinction}
\end{figure}

We now check to see whether the apparent stream could be an artifact of high or low interstellar extinction.  Figure \ref{fig:extinction} shows the extinction map in the sky region around the stream. The white arrows have been included to indicate the apparent position of the structure from Figure \ref{fig:scatter_projection}. From this we conclude that position of the stream does not correlate with any specific structure in the extinction map.

In order to estimate the significance of the detected signal, we selected a region of 333.5 deg$^2$ around the overdensity. This region was then rotated into a coordinate system in which the stream was aligned with the $l'$ axis using the following equations:

\begin{equation}
\begin{split}
\begin{bmatrix}
l'     \\
b'     \\
            \end{bmatrix}
            =
\begin{bmatrix}
\cos (\theta)       &   \sin (\theta)      \\
-\sin (\theta)      &   \cos (\theta)       \\
            \end{bmatrix}
            \cdot
\begin{bmatrix}
(l-l_0)\cos(b)     \\
(b-b_0)    \\
            \end{bmatrix}
+
\begin{bmatrix}
l_0    \\
b_0     \\
            \end{bmatrix}
\end{split}
\end{equation}

\begin{figure}
\includegraphics[width=\columnwidth]{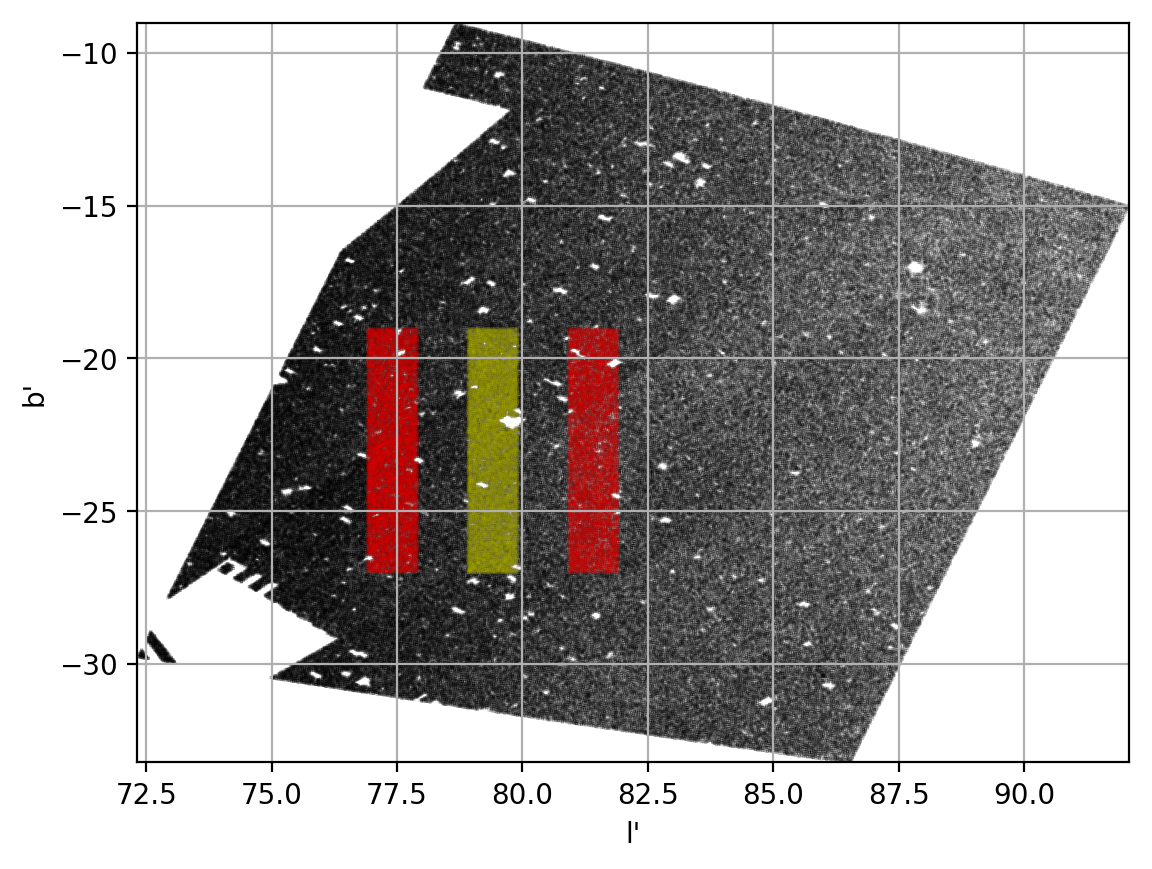}
    \caption{Stellar projection of a smaller region of Figure 2 in the $(l',b')$ coordinate system. The yellow rectangle corresponds to the ``OnStream" Field ($-27\degr < b' < -19\degr$; $78.9\degr < l' < 79.9\degr$). The red rectangles define the ``OffStream" Left Field and ``OffStream" Right Field that are located at ($-27\degr < b' < -19\degr$; $76.9\degr < l' < 77.9\degr$) and ($-27\degr < b' < -19\degr$; $80.9\degr < l' < 81.9\degr$), respectively.}
    \label{fig:stream_on_off}
\end{figure}

The $(l,b)$ coordinate system was rotated around the point $(l_0,b_0) = (79.^{\circ}37,-24.^{\circ}65)$, adopting $\theta = -73^{\circ}$. The longitude was corrected for the change in scale of the latitude. By doing this, the structure now has a vertical orientation in the new ($l'$,$b'$) coordinate system, aligned with $l'\sim 79^{\circ}.4$. Figure \ref{fig:stream_on_off} shows the rotated region around the overdensity.

To confirm that our classic 2D rotation does not significantly distort the area at the stream location, similarly to \citet{shipp2018}, we aligned the stream with the fundamental plane of an arbitrary reference frame whose pole is located at $RA = 226.16^{\circ}$ and $DEC = 28.96^{\circ}$. We represent the longitude and latitude in these coordinates as ($\phi_1$;$\phi_2$). The longitude ($\phi_1$), along the stream, and latitude ($\phi_2$), perpendicular to the stream, of this frame has its origin close to the middle of the stream. The difference in area between the 3D rotation and the 2D rotation is smaller than 1$\%$. In Appendix A we provide the transformation matrix from ICRS to the stream frame.

\begin{figure}
	\includegraphics[width=\columnwidth]{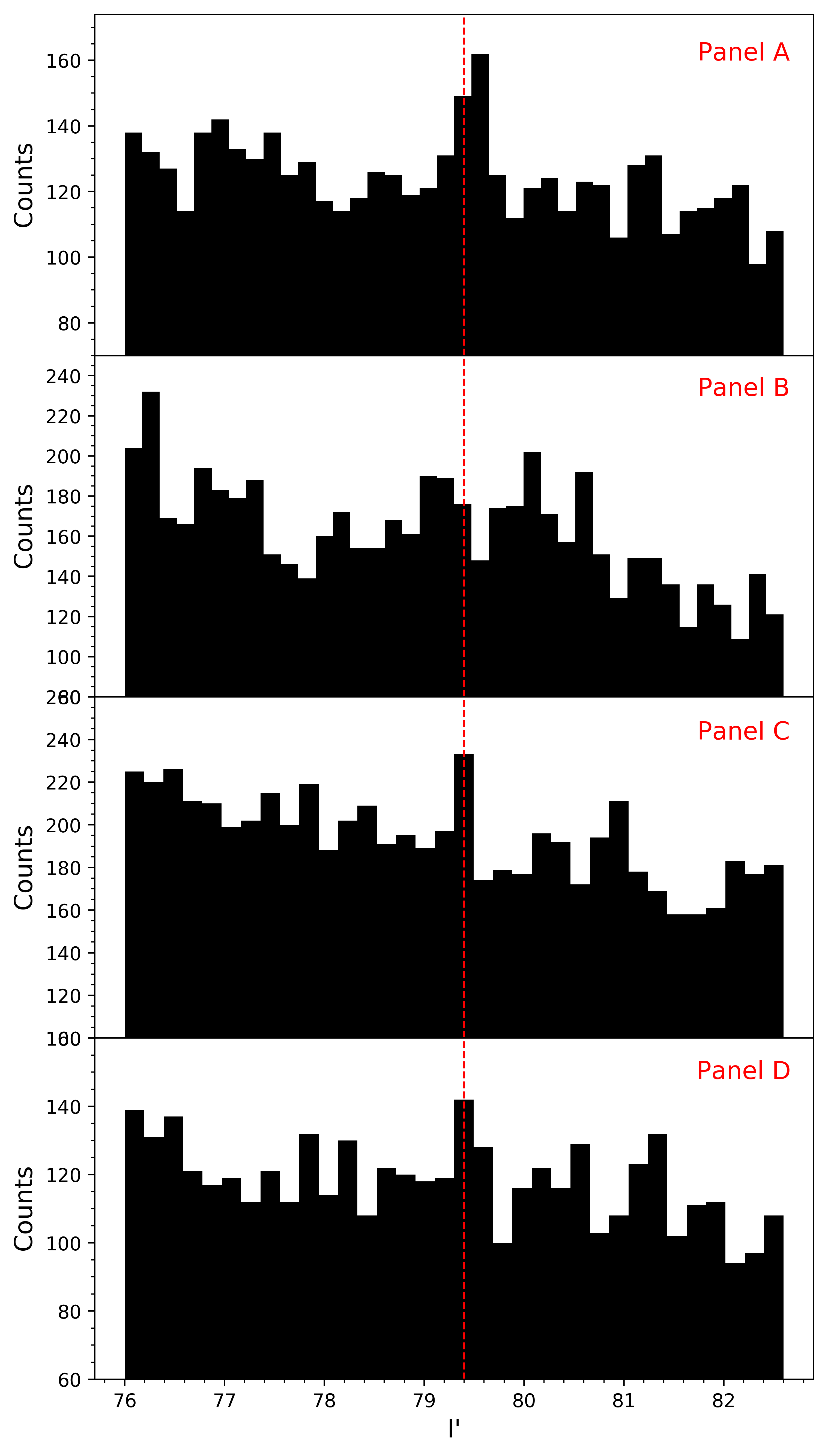}
    \caption{One-dimensional distribution of stellar density integrated between $-27\degr < b' < -19\degr$. Panel A was generated from the SDSSgrgi sample. It is possible to identify a peak at $79.\degr3 < l' < 79.\degr7$. In Panel B, showing the stellar density of faint red dwarfs (the SDSSdwarf sample), a similar peak cannot be found. The stream signature is fainter in the Pan-STARRs data (Panel C, corresponding to the Pan-STARRsgi sample) but still consistent with the results from the SDSS sample; the smaller significance is probably due to the brighter limiting magnitude in Pan-STARRs compared to the SDSS. In fact, when we apply the same magnitude limits of the Pan-STARRsgi sample to the SDSSgrgi sample (Panel D), a very similar result is obtained. In all panels, the red line at $l'\sim 79.\degr4$ marks the position of the stream. Stars in the samples in panels A, B, and D were selected to have a probability > 0.8, using the matched filter. }
    \label{fig:stream_hist}
\end{figure}

We rotated the SDSSgrgi sample to this new coordinate system and used it to generate a 1D profile of the stellar density in the region delimited by $-27\degr < b' < -19\degr$ and $76\degr < l' < 82.6\degr$, shown in Figure \ref{fig:stream_hist}A. This is the same sample used to generate Fig. \ref{fig:scatter_projection}C which gave us the best contrast of stars for the stream. In this panel we can see a clear peak at $l' \sim 79.\degr4$. From this signal, we define an `OnStream' Field ranging from $-27\degr < b' < -19\degr$ and $78.9\degr < l' < 79.9\degr$, an 'OffStream' Left Field, and an `OffStream' Right Field located on either side of the overdensity at ($-27\degr < b' < -19\degr$; $76.9\degr < l' < 77.9\degr$) and ($-27\degr < b' < -19\degr$; $80.9\degr < l' < 81.9\degr$), respectively, as shown in Figure \ref{fig:stream_on_off}. Using the highest stream signal bin of 162 counts, we compared to the average background between 77 and 82 degrees, 120 stars. Using Poisson noise of the background, we estimate the significance of the stream to be 3.83 sigma.

To look for data anomalies that could give us the appearance of a faint stream at this location, we generated a histogram of nearby red dwarfs having the same  magnitude range as that of the sample SDSSgrgi (see Fig. \ref{fig:stream_hist}B). This histogram does not show a peak in the region of the stream. The fact that we cannot identify a similar peak among redder stars in the same magnitude range reinforces the idea that the overdense region identified as a stream is not an artifact of completeness.

We also searched for the signal of the candidate stream in other photometric surveys. Of the public photometric surveys presently available in the literature, only Pan-STARRs has coverage of the same area in both galactic coordinates and magnitude (although it is somewhat shallower than SDSS, going down to $r_0 \approx 21.4$). We applied a similar procedure as was used for the SDSS data to Pan-STARRs, as described in Section 2: in this case, the likelihood filter was applied to the $((g-i)_0,g_0)$ CMD, yielding the sample Pan-STARRsgi (see Table \ref{tab:selections}). Fig. \ref{fig:stream_hist}C shows the histogram of this Pan-STARRsgi sample rotated to the $(l',b')$ system and limited by $-27\degr < b' < -19\degr$ and $76\degr < l' < 82.\degr6$. A peak coincident in $l'$ with that of Fig. \ref{fig:stream_hist}A is clearly seen, showing the same structure is present in the Pan-STARRs data. The ratio of counts in this peak over the adjacent background is slightly smaller than that in Fig. \ref{fig:stream_hist}A, yeilding a significance of $\sim 3.36\sigma$. We believe this is due to Pan-STARRs being shallower than SDSS. To test this, we built a new SDSS sample having the same magnitude limits and using the same likelihood filter used for Pan-STARRsgi. Its histogram is shown Fig. \ref{fig:stream_hist}D; the peak corresponding to the stream is still seen at $l' \sim 79.\degr4$, having an excess over the background very similar to that seen for the Pan-STARRsgi sample. We conclude that the stream we are reporting here is also present in the Pan-STARRs survey.

Other signatures of this structure could not be found, probably due to its faintness. We looked for SDSS stellar spectra but no data seems to cover the extent of the stream. We also searched for its candidate giants stars in LAMOST data using CMDs with JHK magnitudes in agreement with that of the isochrone used in Section 2, but no star passed the selection criteria. We also did a cross-match with the Gaia DR2 data, but the matched stars have large errors in both proper motion and parallax; thus, the analysis with Gaia was inconclusive. It is likely that one or two more years of Gaia observations will be needed before we have useful proper motions for these faint stars.

\begin{figure*}
\begin{center}
  \resizebox{\hsize}{!}{\includegraphics{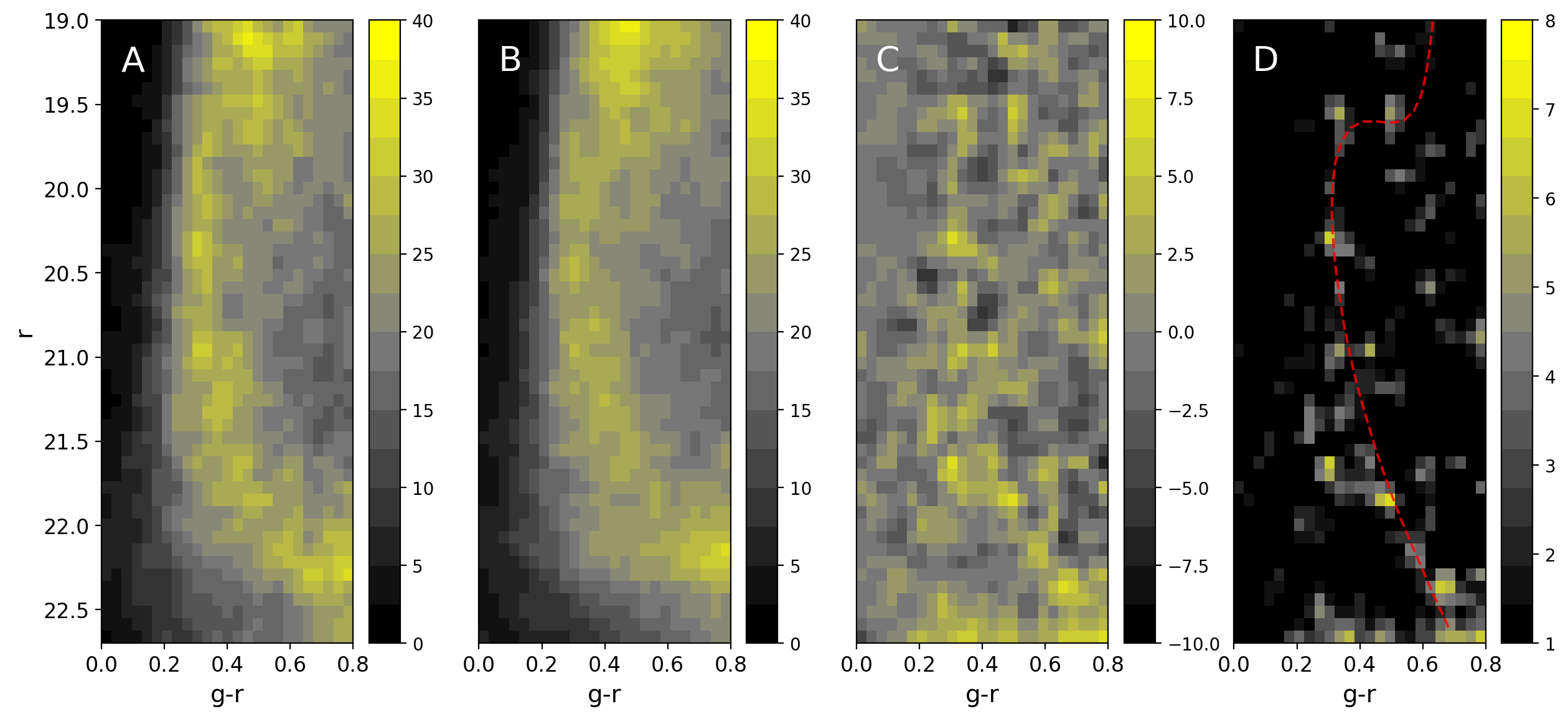}}  
\caption{Panels A and B show the CMD of the OnStream Field and of the OffStream Fields (see Figure \ref{fig:stream_on_off}), respectively. Panel C shows the residual CMD of (OnStream Field $-$ OffStream Fields). Positive numbers indicates bins where OnStream Field has more counts than OffStream Fields. The yellow region of residual counts at $g-r \sim 0.4$ and between $19.5 < r < 21$ is part of the MS of the stream. The MS can be better identified with just the positive residual counts of the residual CMD of (OnStream Field $-$ OffStream Fields), in Panel D. The positive residual counts of the stream can be compared with the isochrone (SSP) that was used at sample selection, marked in red. The Hess diagrams in Panels C and D were
smoothed with a 2D-Gaussian kernel with a standard deviation of 0.75 pixels, as in \citet{shipp2018}.
}
    \label{fig:stream_cmd}
\end{center}
\end{figure*}

\begin{figure*}
	\includegraphics[width=\hsize]{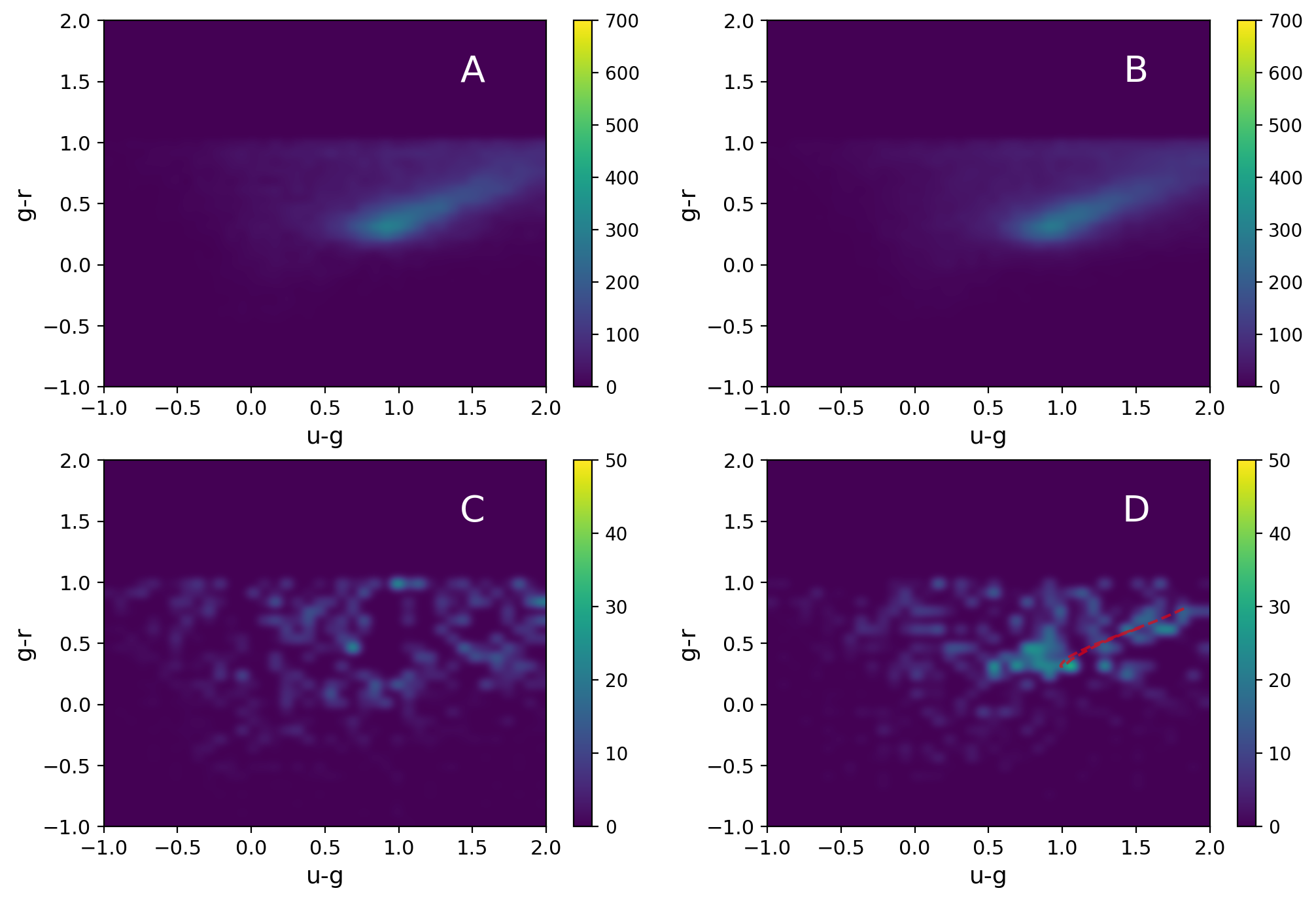}
    \caption{Panels A and B show the color-color magnitudes of the OnStream Field and of the OffStream Fields, respectively.  The overdensity between $1.0<(u-g)_0<2.0$ and $0.3<(g-r)_0<0.8$ is mainly from thick disk and halo stars. Panel C shows the residual color-color diagram of (OffStream Field $-$ OnStream Fields). Panel D shows the opposite residual (OnStream Field $-$ OffStream Fields). The SSP is represented by the red dashed line at Panel D. There is an overdense region at $((u-g)_0, (g-r))_0 \sim (1.07, 0.32)$, which we identify as the stream signal. Using Equation 4 from \citet{ivezic08}, we estimate for the stream a metallicity [Fe/H] $= -$0.75. }
    \label{fig:stream_color_color}
\end{figure*}

Figure \ref{fig:stream_cmd}A shows the CMD for the OnStream Field. The average CMD of the OffStream Left Field and OffStream Right Field (hereafter OffStream Fields) is shown in Figure \ref{fig:stream_cmd}B. Figure \ref{fig:stream_cmd}C shows the residual map subtracting the OffStream Fields from the OnStream Field. In this panel, positive counts (yellow) correspond to an excess in the OnStream field. From panel 6C an excess consistent with a MS for a single stellar population can be seen, particularly for $r_0 > 21.0$. In Figure \ref{fig:stream_cmd}D we show the positive counts from panel 6C to highlight the MS. Additionally, we plot the selection isochrone used for the likelihood filter in Figure \ref{fig:CMD_SDSS_stars_1} is marked in red, which matches the observed excess. In Figure \ref{fig:stream_cmd}C and D, the Hess diagrams have been smoothed using a Gaussian kernel as described in \citet{shipp2018}.

We estimated the metallicity of the stream using a photometric metallicity estimator (\citealt{ivezic08}, \citealt{koposov10}). Figure \ref{fig:stream_color_color} shows four color--color diagrams. Again we have broken the panels down into the OnStream Field, OffStream Fields, and the residuals. Panel A of Figure \ref{fig:stream_color_color} shows ($g-r)_0$ versus ($u-g)_0$ colors of the stars from the OnStream Field, whereas Panel B shows the average color-color diagram from the OffStream Fields. We truncated the sample to $(g-r)_0 < 1.0$ to remove most of contribution of thick disk and halo stars in the color-color diagram. In the residual (OffStream Fields $-$ OnStream Field) shown in Panel C, it is not possible identify any high peak of counts, but only small density fluctuations over all the color-color diagram. However, in Panel D (OnStream Field $-$ OffStream Fields), the positive residual counts appear to have a higher density point at $((u-g)_0, (g-r)_0) \sim (1.07, 0.32)$ which we are attributing to the stream. In Panel D, the SSP is shown as a red dashed line. The peak of this region has more than 40 stars per bin while all the other dense region have at most 27 stars per bin. If we use the equations from \citet{ivezic08} for the peak observed in the residual, we find a metallicity for the stream of [Fe/H] $\sim -0.75$. The 0.3 dex difference in metallicity between isochrone and that from color--color diagram is similar to that from \citet{koposov10}; it is due to the inaccuracy of the color--color diagram method, and would mainly just change the age of the best fit isochrone.

We obtained the luminosity of the stream using the same method of \citet{perottoni18} and PARSEC luminosity functions for a given SSP \citep{bressan12}. The residual magnitude distribution was found by subtracting the OffStream Fields from OnStream Fields only for stars selected to have a probability > 0.8. We estimated the luminosity of the candidate stream as 1-3$\times 10^3 L_\odot$. This indicates that a possible progenitor of the stream is a globular cluster. 

The luminosity function fitted to our residual magnitude distribution indicates that the stream is expected to contain $\sim$ 10 red clump stars over OnStream Field. This could be the reason why no giants for this area and distance were found in the LAMOST data.

\section{Discussion and Conclusions}

We have detected the signature of a stream in the direction of the constellation Pegasus, located at $(l,b) \sim (79.\degr37,-24.\degr65)$. The stream was found using a technique that selects stars with higher probability of belonging to a stellar population in two different color--magnitude diagrams. We show in Figure \ref{fig:scatter_projection} the selection on multiple colors and magnitudes could be used to enhance the signal of the stellar populations as suggested by \citet{perottoni18}.

The theoretical isochrone fit to the stream has an age of 8 Gyr with a metallicity of [Fe/H] $= −0.46$ dex. It is located at a distance of approximately 18 kpc from the Sun (matched filters at 15 kpc and 21 kpc produced no stream signal). The metallicity from the adopted isochrone is compatible with the metallicity obtained from the color-color diagram, indicating that the population of the stream is possibly more metal rich than the other already known streams (e.g. \citealt{Newberg2016, shipp2018}).

The Ivezic method and the metallicity from the isochrones used to extract this SSP seem to have good agreement. However, we acknowledge that there are certain uncertainties to both of these methods and spectroscopic data/members would be needed to confirm this high metallicity stream.

Due to the low surface brightness and narrow width of the stream, the likely progenitor is a globular cluster. Due to a lack of spectroscopic data, we are not able to fit an orbit for this stream, nor link the stream to any other known clusters or stellar substructure.

\section*{Acknowledgements}

HDP, HJR-P and FA-F thank the Brazilian Agency CAPES (DS and PDSE programs) for the financial support of this research. Funding was also provided by the US National Science Foundation grant AST 16-15688, The Marvin Clan, Babette Josephs, Manit Limlamai, and the 2015 Crowd Funding Campaign to Support Milky Way Research, to CM and HJN.

ARGJ thanks FAPESP proc. 2018/11239-8

Funding for SDSS-III has been provided by the Alfred P. Sloan Foundation, the Participating Institutions, the National Science Foundation, and the U.S. Department of Energy Office of Science. The SDSS-III web site is \url{http://www.sdss3.org/}.

SDSS-III is managed by the Astrophysical Research Consortium for the Participating Institutions of the SDSS-III Collaboration including the University of Arizona, the Brazilian Participation Group, Brookhaven National Laboratory, Carnegie Mellon University, University of Florida, the French Participation Group, the German Participation Group, Harvard University, the Instituto de Astrofisica de Canarias, the Michigan State/Notre Dame/JINA Participation Group, Johns Hopkins University, Lawrence Berkeley National Laboratory, Max Planck Institute for Astrophysics, Max Planck Institute for Extraterrestrial Physics, New Mexico State University, New York University, Ohio State University, Pennsylvania State University, University of Portsmouth, Princeton University, the Spanish Participation Group, University of Tokyo, University of Utah, Vanderbilt University, University of Virginia, University of Washington, and Yale University.

The Pan-STARRS1 Surveys (PS1) have been made possible through contributions by the Institute for Astronomy, the University of Hawaii, the Pan-STARRS Project Office, the Max-Planck Society and its participating institutes, the Max Planck Institute for Astronomy, Heidelberg and the Max Planck Institute for Extraterrestrial Physics, Garching, The Johns Hopkins University, Durham University, the University of Edinburgh, the Queen's University Belfast, the Harvard-Smithsonian Center for Astrophysics, the Las Cumbres Observatory Global Telescope Network Incorporated, the National Central University of Taiwan, the Space Telescope Science Institute, and the National Aeronautics and Space Administration under Grant No. NNX08AR22G issued through the Planetary Science Division of the NASA Science Mission Directorate, the National Science Foundation Grant No. AST-1238877, the University of Maryland, Eotvos Lorand University (ELTE), and the Los Alamos National Laboratory.





\appendix

\section{Transformation from equatorial coordinates to Pegasus stream coordinates} \label{sec:coo_appendix}

The transformation matrix is represented as

\begin{eqnarray}
\begin{bmatrix}
\cos(\phi_2) \cos(\phi_1)\\
\cos(\phi_2) \sin(\phi_1)\\
\sin(\phi_2)
\end{bmatrix}&=&\nonumber\\
\begin{bmatrix}
 0.79396012 & -0.44251763 &  0.41689983\\
-0.04887407 &  0.63704904 &  0.76927229\\ 
-0.60600218 & -0.63114711 &  0.4841639
\end{bmatrix} 
&\times &
\begin{bmatrix}
\cos(\alpha) \cos(\delta)\\
\sin(\alpha) \cos(\delta)\\
\sin(\delta)
\end{bmatrix}\nonumber
\end{eqnarray}


\bsp	
\label{lastpage}
\end{document}